\documentstyle[twocolumn,aps,epsfig,subeqnar,pra]{revtex}

\begin{document}

\title{ Bose-Einstein condensates in a one-dimensional double square well: Analytical solutions of the Nonlinear Schr\"odinger equation}

\author{K. W. Mahmud$^{1}$\cite{byline}, J. N. Kutz$^{2}$, and  W. P. Reinhardt$^{1,3}$\\}
\address{$^{1}$Department of Physics, University of Washington, Seattle, WA 98195-1560, USA\\}
\address{$^{2}$Department of Applied Mathematics, University of Washington, Seattle, WA 98195-2420\\}
\address{$^{3}$Department of Chemistry, University of Washington, Seattle, WA 98195-1700, USA\\}

\date{\today}

\maketitle

\begin{abstract}

We present a representative set of analytic stationary state solutions of the Nonlinear Schr\"odinger equation for a symmetric double square well potential for both attractive and repulsive nonlinearity. In addition to the usual symmetry preserving even and odd states, nonlinearity introduces quite exotic symmetry breaking solutions - among them are trains of solitons with different number and sizes of density lumps in the two wells. We use the symmetry breaking localized solutions to form macroscopic quantum superposition states and explore a simple model for the exponentially small tunneling splitting.

\end{abstract}

\section{Introduction}

Many features of Bose-Einstein condensates (BECs) of dilute atomic gases in a single well external potential at zero temperature are well described by mean field theory~\cite{dalfovo1,legget1}. In the  mean field picture all condensate atoms have the same macroscopic wave function satisfying the Gross-Pitaevskii (GP) equation. In this paper we investigate the stationary states of BEC in a symmetric double square well potential. We find analytic solutions of the GP equation for both symmetry preserving and symmetry breaking stationary states of the attractive and repulsive nonlinearity. The solutions presented in the paper give such analytic expressions for what are seen to be stationary soliton trains in the double well - among them are such trains with different number and sizes of density lumps in the two wells. Single dark solitons~\cite{billsoliton,darksoliton2}, bright soliton~\cite{brsolscience} and soliton trains~\cite{brsolnature} have been recently experimentally observed in trapped BECs, suggesting that their double well analogs may be experimentally accessible. In addition we present, as an application of the mean field symmetry breaking solutions, a zero order macroscopic mean field descriptions of macroscopic quantum superposition states (Schr\"odinger Cat state) in a double well BEC system.

Symmetry breaking mean field solutions, such as we observe in this exact treatments, are expected in the attractive case as an attractive condensate in the ground state tends to localize in one well or the other. Symmetry breaking solutions for a nonlinear Schr\"odinger equation was first pointed out in the context of molecular states~\cite{davies}. Symmetry breaking mean field states for repulsive condensates have been discussed in the two-state model of condensate dynamics in a double well~\cite{smerzi1,milburn1,kivshar1,cirac1}, and seen in the nonlinear numerical studies of the GP equation in a symmetric quartic double well~\cite{presilla1}. The present analytic work thus confirms the numerical work of D'Agosta and Presilla in Ref.~\cite{presilla1}  in the context of a double square well. Such macroscopic quantum self-trapped states have also appeared on the studies of transport on a dimer modeled by discrete nonlinear Schr\"odinger equation~\cite{eilbeck}.

BECs in a double well and multi-well systems have been studied in the context of coherence~\cite{ketterle2}, Josepson tunneling~\cite{smerzi1,javanainen1,inguscio1}, squeezed states~\cite{kasevich2}, the superfluid to Mott transition~\cite{mott1} and condensate fragmentation~\cite{spekkens1}. In discussions of condensate tunneling it is well known that a high barrier leads to condensate fragmentation in which two or more distinct single particle states are macroscopically occupied. For a repulsive condensate, raising the barrier leads to the condensate in the two wells from being coherent to being incoherent in a Fock state~\cite{spekkens1}. The analysis herein gives the nonlinear modes of the entire double well in a mean field picture when all the atoms have the same single particle wavefunction. Correlation effects leading to condensate fragmentation are neglected here and thus the theory presented applies directly only to the case of strong tunneling. However, the mean field states obtained could form the basis for a correlated description.

The GP equation is a cubic nonlinear Schr\"odinger equation(NLSE)~\cite{sulem1} where the particle interactions give rise to such effective nonlinearity. The NLSE has been successful in modeling many other natural phenomenon besides BEC. It describes light pulses in optical fibers~\cite{hasegawa1}, helical excitations of a vortex line~\cite{hasimoto1}, Bose-condensed photons~\cite{ciao1}, spin waves in magnetic materials~\cite{kalinikos1}, and disordered media~\cite{mamaev1}. Despite being a canonical physics problem~\cite{cohenT1}, the symmetric double square well problem has not, to our knowledge, been solved for nonlinear Schr\"odinger equation. Although the discussions in the paper is exclusively for Bose-Einstein condensates, the analysis will apply to any system satisfying cubic NLSE. 

The symmetry breaking localized one particle mean field states can be used to form a zero order two-configuration Schr\"odinger cat states of the form $\phi_{left}^N \pm \phi_{right}^N$. There have been several reports of the creation of Schr\"odinger cat states in various condensed matter systems~\cite{friedman1,cat2}. In the context of BEC, several authors have suggested producing such states~\cite{cirac1,savage1,burnett1,ruostekoski1,ruostekoski2}, although none have been demonstrated experimentally. In a double well, as is found analytically in this paper, the mean field ground state for an attactive condensate is a symmetry breaking state localized in one of the wells. The superposition of such degenerate localized states is a ``cat" state. We calculate the tunneling splittings for such states using correct mean field single particle states starting from the full N-body Hamiltonian. Such two-configuration tunneling splittings are exponentially small in the N-body wave function overlap. Particle correlations are still neglected, but strong mean field effects accounted for. 

The article is organized as follows. In Sec.~\ref{sec:dwell} we present the full set of symmetry preserving and symmetry breaking analytic solutions of stationary NLSE for a symmetric double square well potential. In Sec.~\ref{sec:cat} we discuss an application of the symmetry breaking solutions - the possibility of creating superpositions of macroscopic quantum states, and calculate the tunneling splittings of such cat states. Remarks and discussions in Sec.~\ref{sec:summary} conclude the paper.

\section{Double Square Well}
\label{sec:dwell}

The stationary NLSE with a potential has the form 
\begin{equation}
[-\partial_{x}^{2}+\eta\mid\!f(x)\!\mid^{2}+V^{trap}(x)\,]\,f(x)=\mu\,f(x)\,,
\label{eqn:nls}
\end{equation}
where $f(x)$ is the mean field condensate wavefunction in the longitudinal direction, $\mu$ is the eigenvalue or the chemical potential, and $\eta$ is the nonlinearity parameter which is proportional to the number of atoms and the s-wave scattering length. All quantities in Eq.(\ref{eqn:nls}) are dimensionless.

\begin{figure}
\begin{center}
\epsfig{figure=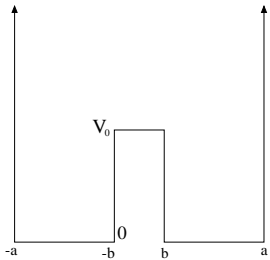,width=3.5cm}
\end{center}
\caption{Symmetric double square well potential: the model used in this paper}
\label{fig:potential}
\end{figure}

Analytic solutions of the GP equation for harmonic and quartic double well potentials are not possible, so we have chosen to investigate the infinite square well with symmetrically placed finite rectangular potential barrier. The potential is of the form 
\begin{equation}
V^{trap}(x)=\left\{\begin{array}{lll}
    \infty, \quad & |x|\geq a\\
    0, \quad &  b<|x|<a\\
    V_o, \quad & |x|<b
    \end{array}
  \right.
\label{eqn:potential}
\end{equation}
For clarity, Fig.\ref{fig:potential} shows a picture of this potential. Double well traps can be created in experiments with a combination of optical and magnetic trapping. Varying the laser strengths the barrier or the depth and the width of the trap can be easily tailored to experimental specifications. The double well traps created in experiments usually have gaussian barriers; however, the qualitative behavior of the stationary states of such wells would be the same as discussed in this paper for a double square well.

We present the analytic solutions of Eq.(\ref{eqn:nls}) with the potential Eq.(\ref{eqn:potential}). Solutions in an infinite well and a finite well have been presented for both attractive and repulsive condensates~\cite{lcarr2,lcarr1,kmahmud1}. In Eq.(\ref{eqn:nls}) $\eta>0$ corresponds to repulsive condensate while $\eta<0$ corresponds to attractive condensate. The solutions of NLSE in a zero potential are Jacobian Elliptic functions~\cite{stegun1}. Such functions are well known in the soliton literature, and also as the solution to the anharmonic classical oscillator, i.e. $\ddot{\theta}+\theta-\theta^{3}/3!=0$. An example of the standard notation for Jacobi Elliptic functions is sn$(x\mid m)$, where $m$ is the elliptic parameter. The period is given by $4 K(m)$, where $K(m)$ is the complete elliptic integral. The value of $m$ is bounded between $0$ and $1$. It interpolates the elliptic functions between trigonometric and hyperbolic functions. There are 12 elliptic functions all of which are solutions to the NLSE. Of the 12 elliptic functions, six are bounded and six are unbounded. Of the six bounded functions, only sn$(x\mid m)$, cn$(x\mid m)$, dn$(x\mid m)$ have distinct physical forms. Others differ only by a translational shift or a rescaling of the amplitude. The six unbounded functions can be represented as a quotient of the above three functions in different combinations. We will find that the pieces of these unbounded functions are those appropriate in the barrier region of the double well for a repulsive condensate. Table~\ref{table:jacobian} summarizes the functions relevant to this work.

Solutions in the three regions will be written in the form
\begin{equation}
f(x)=\left\{\begin{array}{lll} 
   f_1(x),\quad & -a<x<-b \\
   f_2(x),\quad & |x|\leq b\\
   f_3(x),\quad & b<x<a
  \end{array}
 \right.
\label{eqn:form}
\end{equation}
The solutions vanish on and outside the hard wall boundary at $|x|\geq a$. The solutions will be found subject to continuity of $f(x)$ and $f^{\prime}(x)$ at $x=\pm b$ and the normalization condition $\int_{-a}^{a}dx\,|f(x)|^2=1$. The vanishing of the solutions at the hard walls is taken as built into the elliptic functions and does not form an additional boundary condition. The solutions are divided into two different categories - Symmetry preserving and Symmetry breaking. Taking advantage of the symmetry of the problem finding symmetry preserving states reduce to solving a set of three nonliear algebraic equations. The symmetry breaking states require solving five simultaneous nonlinear equations which is a far more difficult undertaking.   

\subsection{Symmetry preserving states}

Symmetry preserving states are the states that preserve the symmetry of the N-particle many-body Hamiltonian. Simply put, they are the even and odd solutions. As we will find out in the next section, there can also be solutions which does not preserve even/odd symmetry expected from linear quantum mechanics.

\subsubsection{Attractive nonlinearity}
Symmetric solutions take the following form  
\begin{subeqnarray}
\label{eqn:spA1}
  f_1(x)&=& A\,\text{cn}(k (x+a)-K(m)\mid m)\, ,\\
  f_2(x)&=& A_2\,\text{dn}(k_2 x+K(m_2)\mid m_2)\, ,\\
  f_3(x)&=& A\,\text{cn}(k (x-a)+K(m)\mid m)\,,
\end{subeqnarray}
and antisymmetric solutions take the form
\begin{subeqnarray}
\label{eqn:spA2}
 f_1(x)&=& A\,\text{cn}(k(x+a)-K(m)\mid m)\, ,\\
 f_2(x)&=& A_2\,\text{cn}(k_2 x+K(m_2)\mid m_2)\,,\\
 f_3(x)&=& -A\,\text{cn}(k(x-a)+K(m)\mid m)\,,
\end{subeqnarray}

where $A$, $A_2$, $k$, $k_2$, $m$ and $m_2$ are free parameters. $f_1(x)$ and $f_3(x)$ have been chosen to preserve odd and even parity. Note that the elliptic parameter $K(m_2)$ displaces the $cn$ in the barrier region to make it antisymmetric. In the next section we describe uniquely nonlinear type solutions which does not preserve such parity. The condition that the states vanish at the hard walls at $a$ and -$a$ are built into the form of the solutions.

Symmetric and antisymmetric solutions are solved using the same method. Substituting the symmetric solutions into Eq.~(\ref{eqn:nls}) with the potential Eq.~(\ref{eqn:potential}), following conditions are obtained
\begin{subeqnarray}
\label{eqn:spA3}
 A^2&=& 2mk^2/\eta,\, A_2^2=2k_2^2/\eta\\
\mu&=& (1-2m) k^2,\, \mu=(m_2-2)k_2^2+V_o
\end{subeqnarray}
The boundary condition $f_1(-b)=f_2(-b)$ is equivalent to $f_2(b)=f_3(b)$, and requires
\begin{equation}
\label{eqn:spA4}
A\,\text{cn}(k \omega -K(m)\mid m)=A_2\,\text{dn}(-k_2 b+K(m_2)\mid m_2)
\end{equation}
where $\omega \equiv a-b$ is the width of each of the wells.
Continuity of the first derivative requires
\begin{eqnarray}
\label{eqn:spA5}
&A k\,\text{sn}(k \omega-K(m)\mid m)\,\text{dn}(k \omega-K(m)\mid m)=A_2 m_2 k_2 \nonumber\\
&\times \text{sn}(-k_2 b +K(m_2)\mid m_2)\,\text{cn}(-k_2 b+K(m_2)\mid m_2)
\end{eqnarray}
Finally, the normalization condition is
\begin{eqnarray}
\label{eqn:spA6}
2A^2\int_b^a dx\,\text{cn}^2(k(x-a)+K(m)\mid m)\nonumber\\
+2A_2^2\int_0^b dx\, \text{dn}^2 (k_2 x+K(m_2)\mid m_2)=&1\,.
\end{eqnarray}
Eq.~(\ref{eqn:spA6}) can be written as
\begin{eqnarray}
\label{eqn:spA7}
&\frac{2 A_2^2}{k_2}\,[E(k_2 b+K(m_2)\mid m_2)-E(m_2)]-\frac{2A^2}{m} (1-m) \omega\nonumber\\
& + \frac{2A^2}{m k} [E(m)-E(-k \omega+K(m)\mid m)]=1\,.
\end{eqnarray}
where $E(k_2 l\mid m)$ is standard notation for an incomplete elliptic integral~\cite{stegun1}.

Equating of Eqs.~(\ref{eqn:spA3}b) gives us a constraint on the energy. Substitution of Eqs.~(\ref{eqn:spA3}a) into Eqs.~(\ref{eqn:spA4}), ~(\ref{eqn:spA5}), and ~(\ref{eqn:spA7}) produces a system of four simultaneous equations - an energy condition, a nontrivial normalization and two enforcing the continuity of the wavefunction and its first derivative at the interior discontinuity of the potential. The four equations can be reduced to three equations in three unknowns. These are
\begin{subeqnarray}
\label{eqn:spA8}
&\sqrt{m} k\,\text{cn}(k \omega-K(m)\mid m)\nonumber\\
&= \lambda\,\text{dn}(-\lambda b+K(m_2)\mid m_2)\\
&\nonumber\\
&\sqrt{m} k^2\,\text{sn}(k \omega -K(m)\mid m)\,\text{dn}(k \omega -K(m)\mid m)= m_2\lambda^2 \nonumber\\
&\text{sn}(-\lambda b+K(m_2)\mid m_2)\,\text{cn}(-\lambda b+K(m_2)\mid m_2)\\
&\nonumber\\
&\frac{4 \lambda}{\eta}\,[E(\lambda b+K(m_2)\mid m_2)-E(m_2)]-\frac{4 k^2}{\eta}\,(1-m)\omega\,\nonumber\\
&+\frac{4 k}{\eta}[E(m)-E(-k \omega +K(m)\mid m)]=1\,.
\end{subeqnarray}
 where $\lambda=\sqrt{\frac{V_o-(1-2m)k^2}{2-m_2}}\equiv k_2$ and $\omega \equiv a-b$. This is a system of three nonlinear algebraic equations with three unknown variables $m$, $m_2$ and $k$ and four experimental parameters - the box width $2a$, barrier height $V_o$, barrier width $2b$ and nonlinearity parameter $\eta$.

This system of equation Eqs.~(\ref{eqn:spA8}) is analogous to the set of equations for linear Schr\"odinger equation for a particle on a box double well potential~\cite{cohenT1}. However, the normalization equation Eq.~(\ref{eqn:spA8}c) here is nontrivial and gives an additional condition. These equations can  ideally be solved by a multidimensional secant method, and that is the method we use to find the roots. However, the nonlinear parameter space is too large to choose a good starting point for the roots to converge. As we will see in the next section when we deal with a set of five equations for the symmetry breaking solutions, it is almost impossible to find the roots and the analytic solutions without a good initial choice of parameters from an approximate numerical solution. 

Such numerical approximations to the exact solutions of Eq.(1) with the double well potential Eq.(2) can be generated by the shooting method~\cite{keller1}. However, the cubic nonlinearity generated from the mean-field interactions of the atoms introduces numerical stiffness into the resulting two-point boundary value problem.  To accurately compute the numerical solutions, Gear's methods~\cite{gear1} are employed which are efficient in overcoming the numerical stiffness by utilizing backward differencing formulas. The resulting shooting scheme is then easily implemented and both the normalized symmetry preserving and symmetry-breaking states are computed along with their chemical potential.  We note that by adjusting the shooting angle, the normalization to unity can be satisfied.

\begin{figure}
\begin{center}
\epsfig{figure=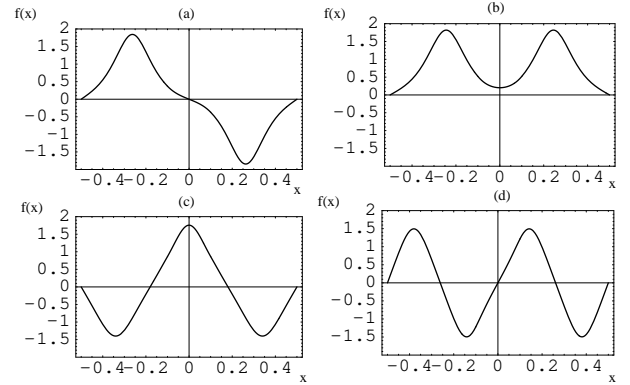,width=8.2cm}
\end{center}
\caption{Shown are the first four symmetry preserving states for attractive nonlinearity. The barrier walls are at x=$\pm$ 0.1}
\label{fig:attrsp}
\end{figure}

Knowing the chemical potential and the value of the solution at barrier boundary $x=b$ from the shooting routine numerics we can find the three approximate roots of Eqs.~(\ref{eqn:spA8}). With the form of the solutions and the approximate roots at hand, secant method is used to solve the Eqs.~(\ref{eqn:spA8}) to find the exact analytic solutions. In Fig.2 we show the first four odd and even states. The states are ordered according to the chemical potential $\mu$. A barrier height of $V_o=100$, barrier width of $2b=1/5$, well width $2a=1$ and nonlinearity of $\eta=-100$ were used. Table~\ref{table:sp} shows the solution parameters for Fig.2. The true mean field ground state for an attractive condensate in this case is a symmetry breaking state where the condensate localizes in one well or the other as is described in the next section. The first excited even state for this well in Fig.~\ref{fig:attrsp}(c) where the condensate has one of the peaks on top of the barrier is a uniquely nonlinear state~\cite{presilla1} which does not have any counterpart in linear Schr\"odinger equation. For $\mu > V_o$ all even solutions are of this kind, however even for $\mu<V_o$ strong nonlinearity can give rise to such states. Symmetric solutions of this kind has the form $f_2(x)= A_2\,\text{cn}(k_2 x\mid m_2)$. 
 
The antisymmetric solutions were found using a similar method. For reference the system of equations is $\sqrt{m_2} \lambda \text{cn}(-\lambda b+K(m_2),m_2)=\sqrt{m} k \text{cn}(k(a-b)-K(m),m)$;\, $\sqrt{m_2} \lambda^2 \text{sn}(-\lambda b+K(m_2),m_2) \text{dn}(-\lambda b+K(m_2),m_2)=\sqrt{m} k^2 \text{sn}(k \omega-K(m),m) \text{dn}(k \omega-K(m),m);\,\frac{4 \lambda}{\eta}\,[E(\lambda b+K(m_2)\mid m_2)-E(m_2)]-\frac{4 k^2}{\eta}\,(1-m)\omega-\frac{4 \lambda^2}{\eta}\,(1-m_2) b+\frac{4 k}{\eta}(E(m)-E(-k \omega +K(m)\mid m))=1$, where $\lambda=\sqrt{\frac{V_o+(1-2 m)k^2}{2m_2-1}}\equiv k_2$.

We would like to note that unlike linear quantum mechanics, for attractive condensate the eigenvalue or chemical potential of the antisymmetric state for this well dimensions has a lower value than the symmetric case. This behaviour is only true for strong nonlinearity. The total energy per particle for the antisymmetric state is however always greater than the symmetric case. Similar behavior of symmetric and antisymmetric state chemical potentials has also been found in the case of ring potentials~\cite{lcarr2}.

\subsubsection{Repulsive nonlinearity}
Symmetric solutions take the form
\begin{subeqnarray}
\label{eqn:spR1}
  f_1(x)&=& A\,\text{sn}(k(x+a)\mid m)\, ,\\
  f_2(x)&=& A_2\,\text{ds}(k_2 x+K(m_2)\mid m_2)\, ,\\
  f_3(x)&=& -A\,\text{sn}(k (x-a)\mid m)\,,
\end{subeqnarray}
and antisymmetric solutions take the form
\begin{subeqnarray}
\label{eqn:spR2}
 f_1(x)&=& A\,\text{sn}(k(x+a)\mid m)\, ,\\
 f_2(x)&=& A_2\,\text{cs}(k_2 x+K(m_2)\mid m_2)\,,\\
 f_3(x)&=& A\,\text{sn}(k(x-a)\mid m)\,,
\end{subeqnarray}
Substitution of theses solutions into Eq.~(\ref{eqn:nls}) with the double well potential Eq.~(\ref{eqn:potential}) gives the following equations for the amplitude and the chemical potential 
\begin{subeqnarray}
\label{eqn:spR3}
 A^2&=& 2mk^2/\eta,\, A_2^2=2k_2^2/\eta\\
\mu&=& (1+m) k^2,\, \mu=-(2 m_2-1)k_2^2+V_o
\end{subeqnarray}
Just like for the attractive case the three simultaneous equations obtained from the boundary conditions, normalization and the energy conditions are following 
\begin{subeqnarray}
\label{eqn:spR4}
&\sqrt{m}\,k\,\text{sn}(k \omega\mid m)=\lambda\,\text{ds}(-\lambda b+K(m_2)\mid m_2)\,,\\
&\nonumber\\
&\sqrt{m}\,k^2\,\text{cn}(k \omega\mid m)\,\text{dn}(k \omega\mid m)=-\lambda^2 \times\nonumber\\
&\text{cs}(-\lambda b+K(m_2)\mid m_2)\,\text{ns}(-\lambda b+K(m_2)\mid m_2)\\
&\nonumber\\
&4 \lambda^2 b/\eta-4\lambda^2 m_2b/\eta+\frac{4 k^2}{\eta}\omega \nonumber\\
&+\frac{2 \lambda}{\eta}[\text{cs}(-\lambda b+K(m_2)\mid m_2)\text{dn}(-\lambda b+K(m_2)\mid m_2)\nonumber\\
&-\text{cs}(\lambda b+K(m_2)\mid m_2)\text{dn}(\lambda b+K(m_2)\mid m_2)]\nonumber\\
&-\frac{2 \lambda}{\eta}[-E(\lambda b+K(m_2)\mid m_2)+E(-\lambda b+K(m_2)\mid m_2)]\nonumber\\
&-\frac{4 k}{\eta}E(k \omega\mid m)=1
\end{subeqnarray}
where $\lambda=\sqrt{\frac{(1+m)k^2-V_o}{1-2 m_2}}\equiv k_2$. A similar set of equations is obtained for the antisymmetric case.

\begin{figure}
\begin{center}
\epsfig{figure=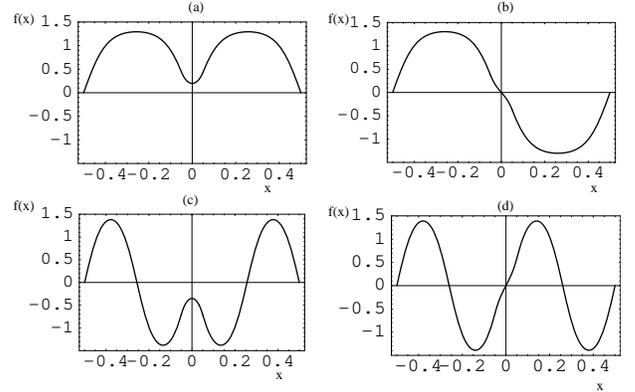,width=8.2cm}
\end{center}
\caption{Shown are the first four symmetry preserving states for repulsive nonlinearity. The barrier walls are at x=$\pm$ 0.05}
\label{fig:repsp}
\end{figure}

The ground state and the first three symmetry preserving excited states are shown in Fig.~(\ref{fig:repsp}). The well dimensions used here are different than the attractive case which was chosen to show the peculiarities of attractive condensate. A barrier height of $V_o=1000$, barrier width of $2b=1/10$, well width $2a=1$ and nonlinearity of $\eta=100$ were used here. Table~\ref{table:sp} shows the solution parameters for Fig.3. In addition to the even and odd excited states there can also be symmetry breaking states as described in the next section. For a repulsive condensate the lowest symmetry preserving state is always the ground state.

\subsection{Symmetry breaking states}

Symmetry breaking states are uniquely nonlinear states where different size or number of ``lumps" are present in the two wells. Such stationary states with strong localization and different number of nodes in the two symmetric wells are not possible for linear Sturm-Liouville systems. Finding such solutions confirms and extends the numerical work~\cite{presilla1} and the two state tunneling models~\cite{smerzi1,milburn1,kivshar1,smerzi2} of the double well where macroscopic quantum self-trapping has been predicted. On the N-particle level the stationary states should preserve the symmetry of the Hamiltonian and can only be symmetric and antisymmetric. So these asymmetric states arise due to the nonlinearity associated with the mean field approximation.

In the work of D'Agosta and Presilla~\cite{presilla1} a non-linear trial function and relaxation method for patial differential equations was used to numerically find both the symmetry preserving and symmetry breaking states of the GP equation in a symmetric harmonic/quartic double well. The difficult task of choosing the right trial functions and the possibility of false minima leading to artifacts in such methods motivated us to treat the model double square well potential and to find the roots of these algebraic equations, and thus find the exact analytic solutions. The qualitative behaviour of solutions in any symmetric double potential should be the same as ours, and wherever the set of parameters we used overlaps with those of Ref~\cite{presilla1} there is one-to-one correspondence in the solutions.

\subsubsection{Attractive nonlinearity}
Solutions with no nodes inside the barrier region take the form
\begin{subeqnarray}
\label{eqn:sbA1}
  f_1(x)&=& A_1\,\text{cn}(k_1 (x+a)-K(m_1)\mid m_1)\, ,\\
  f_2(x)&=& A_2\,\text{dn}(k_2 (x+d)+K(m_2)\mid m_2)\, ,\\
  f_3(x)&=& A_3\,\text{cn}(k_3 (x-a)+K(m_3)\mid m_3)\,,
\end{subeqnarray}
and solutions with nodes inside the barrier are
\begin{subeqnarray}
\label{eqn:sbA2}
 f_1(x)&=& A_1\,\text{cn}(k_1(x+a)-K(m_1)\mid m_1)\, ,\\
 f_2(x)&=& A_2\,\text{cn}(k_2 (x+d)+K(m_2)\mid m_2)\,,\\
 f_3(x)&=& -A_3\,\text{cn}(k_3(x-a)+K(m_3)\mid m_3)\,,
\end{subeqnarray}
$d$ in Eqs.~(\ref{eqn:sbA1}) and ~(\ref{eqn:sbA2}) is a measure of how far the solution under the barrier is displaced from being symmetric. The amplitudes and the chemical potentials are
\begin{subeqnarray}
\label{eqn:sbA3}
 A_1^2&=& 2m_1k_1^2/\eta,\, A_2^2=2k_2^2/\eta,\, A_3^2=2m_3k_3^2/\eta \\
\mu&=& (1-2m_1) k_1^2,\, \mu=(m_2-2)k_2^2+V_o,\nonumber\\
\mu&=&(1-2m_3)k_3^2\,
\end{subeqnarray}
The set of five equations in five unknowns are 
\begin{eqnarray}
\label{eqn:sbA4}
&\sqrt{m_1}\,\alpha\,\text{cn}(\lambda_3(m_1)\mid m_1)=\beta\,\text{dn}(\lambda_1(d,m_2)\mid m_2),\\
&\nonumber\\
&\sqrt{m_3}\,\gamma\,\text{cn}(\lambda_4(m_3)\mid m_3)=\beta\,\text{dn}(\lambda_2(d,m_2)\mid m_2)\,,\\
&\nonumber\\
&\sqrt{m_1}\,\alpha^2\,\text{sn}(\lambda_3(m_1)\mid m_1)\,\text{dn}(\lambda_3(m_1)\mid m_1)\nonumber\\
&=m_2^2 \beta^2\,\text{sn}(\lambda_1(d,m_2)\mid m_2) \text{cn}(\lambda_1(d,m_2)\mid_2)\,,\\
&\nonumber\\
&\sqrt{m_3}\,\gamma^2\,\text{sn}(\lambda_4(m_3)\mid m_3)\,\text{dn}(\lambda_4(m_3)\mid m_3)\nonumber\\
&=m_2^2 \beta^2\,\text{sn}(\lambda_2(d,m_2)\mid m_2)\,\text{cn}(\lambda_2(d,m_2)\mid_2)\,,\\
&\nonumber\\
&-\frac{2\gamma^2}{\eta}(1-m_3)\omega+\frac{2\gamma}{\eta}[E(m_3)-E(\lambda_4(m_3)\mid m_3)]\nonumber\\
&-\frac{2\alpha^2}{\eta}(1-m_1)\omega+\frac{2\alpha}{\eta}[E(m_1)+E(\lambda_3(m_1)\mid m_1)]\nonumber\\
&\frac{2\beta}{\eta}[E(\lambda_2(d,m_2)\mid m_2)-E(\lambda_1(d,m_2)\mid m_2)]=1
\end{eqnarray}
where $\alpha=\sqrt{\frac{\mu}{1-2 m_1}}\equiv k_1$, $\beta=\sqrt{\frac{\mu-V_o}{m_2-2}}\equiv k_2$,
$\gamma=\sqrt{\frac{\mu}{1-2 m_3}}\equiv k_3$, $\lambda_1(d,m_2)=k_2(d-b)+K(m_2)$, $\lambda_2(d,m_2)=k_2(d+b)+K(m_2)$, $\lambda_3(m_1)=\alpha \omega-K(m_1)$ and $\lambda_4(m_3)=-\gamma \omega +K(m_3)$. This is a set of five nonlinear equations in five unknowns $m_1$, $m_2$, $m_3$, $d$ and $\mu$. A similar set of equations are obtained for the solutions that has nodes inside the barrier.
 
As described in the previous section we use a shooting method to find the approximate numerical solutions. Knowing the eigenvalue and the values of the functions at the barrier boundaries at $x=\pm b$, we can reduce five equations with five unknowns to equations with two unknowns. With just two unknowns we can use a graphical method~\cite{kmahmud1} to find the approximate solutions. Such approximate roots are then used to find the exact analytic roots of these five equations using a multidimensional secant method, and thus we obtain the analytic solution of the symmetry breaking states. Without a good bracketing on the roots obtained from first solving it numerically its extremely unlikely for a secant method to converge for a set of five nonlinear equations.     

\begin{figure}
\begin{center}
\epsfig{figure=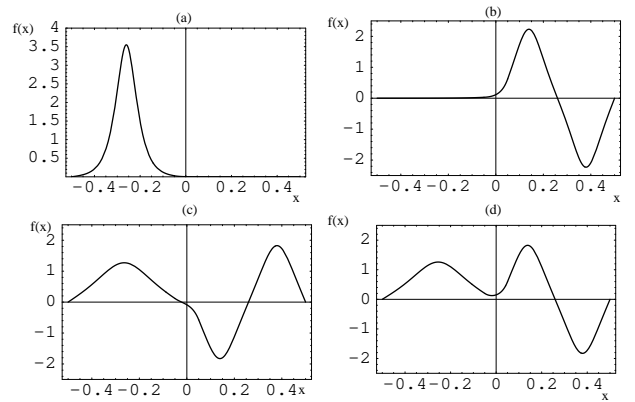,width=8.2cm}
\end{center}
\caption{Shown are the first four zero-node, one-node and two-node  symmetry breaking states for attractive nonlinearity. The barrier walls are at x=$\pm$ 0.05}
\label{fig:attrsb}
\end{figure}

For Fig.~\ref{fig:attrsb} we use a well dimension of $2a=1$, $2b=1/10$, $V_o=1000$ and nonlinearity $\eta=-100$. We use a different well dimension here than the attractive symmetric case just to show the varieties of asymmetric solutions. Table~\ref{table:sb} shows the solution parameters for Fig.4. The solutions can be classified as multiple node solutions - zero node, one node, two node and such. The lowest symmetry breaking state for attractive condensate is the ground state of the system as the clustering of particles in one of the wells minimizes the energy for strong enough self interaction. There can be ground and excited state solutions with assymetrically placed peaks on top of the barrier. The analytic form of such solutions is $f_2(x)= A_2\,\text{cn}(k_2 (x+d)\mid m_2)$. For an asymmetric ground state, increasing the barrier height localizes the condensate more into the well, on the other hand increasing the barrier width pushes the peak of the condensate density more towards the center of the well on top of the barrier. 

\begin{figure}
\begin{center}
\epsfig{figure=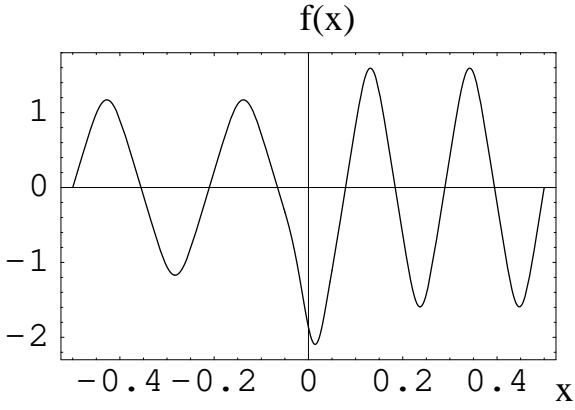,width=8.2cm}
\end{center}
\caption{Shown is a symmetry breaking eight-soliton bright soliton train solution in a double well. The barrier walls are at x=$\pm$ 0.05.}
\label{fig:soltrain}
\end{figure}

Fig.~\ref{fig:attrsp} and Fig.~\ref{fig:attrsb} are the bright soliton solutions in a double well. It shows the one, two, three and four soliton solutions. Bright soliton and soliton trains have recently been observed in attractive condensates of $^{7}$Li~\cite{brsolscience,brsolnature}. Unlike stationary soliton trains of equal density lumps in a single potential well, double well geometry has stationary soliton train solutions with unequal density lumps as is shown in Fig.~\ref{fig:attrsb}. There exists a whole class of such many-soliton solutions. As an example, Fig.~\ref{fig:soltrain} shows an analytic solution of a symmetry breaking eight-soliton bright soliton train in a well of dimensions $2a=1$, $2b=0.1$, $V_o=1000$, and for nonlinearity $\eta=-500$.

\subsubsection{Repulsive nonlinearity}
Solutions with no nodes inside the barrier are
\begin{subeqnarray}
\label{eqn:rsbA3}
  f_1(x)&=& A_1\,\text{sn}(k_1 (x+a)\mid m_1)\, ,\\
  f_2(x)&=& A_2\,\text{ds}(k_2 (x+d)+K(m_2)\mid m_2)\, ,\\
  f_3(x)&=& A_3\,\text{sn}(k_3 (x-a)\mid m_3)\,,
\end{subeqnarray}
Solutions with nodes inside the barrier are
\begin{subeqnarray}
\label{eqn:rsbA4}
 f_1(x)&=& A_1\,\text{sn}(k_1(x+a)\mid m_1)\, ,\\
 f_2(x)&=& A_2\,\text{cs}(k_2 (x+d)+K(m_2)\mid m_2)\,,\\
 f_3(x)&=& -A_3\,\text{sn}(k_3(x-a)\mid m_3)\,,
\end{subeqnarray}
The amplitude and chemical potentials for the states that has no nodes inside the barrier are
\begin{subeqnarray}
\label{eqn:sbr1}
A_1^2&=&2m_2k_1^2/\eta, A_2^2=2k_2^2/\eta, A_3^2=2m_3k_3^2/\eta\\
\mu_1&=&(1+m_1)k_1^2, \mu_2=(1-2m_2)k_2^2+V_o,\nonumber\\ 
\mu_3&=&(1+m_3)k_3^2
\end{subeqnarray}
For reference the equations are
\begin{eqnarray}
\label{eqn:sbr2}
&\sqrt{m_1}\alpha \text{sn}(\alpha\,\omega\mid m_1)=\beta \text{ds}(\lambda_1(d,m_2)\mid m_2)\\
&\nonumber\\
&\sqrt{m_3}\gamma \text{sn}(-\gamma \omega\mid m_3)=\beta \text{ds}(\lambda_2(d,m_2)\mid m_2)\\
&\nonumber\\
&\sqrt{m_1}\alpha^2 \text{cn}(\alpha\omega,m_1) \text{dn}(\alpha\omega,m_1)\nonumber\\
&=-\beta^2 \text{ns}(\lambda_1(d,m_2)\mid m_2)\text{cs}(\lambda_1(d,m_2)\mid m_2)\\
&\nonumber\\
&\sqrt{m_3}\gamma^2 \text{cn}(-\gamma \omega\mid m_3) \text{dn}(-\gamma \omega\mid m_3)\nonumber\\
&=-\beta^2 \text{ns}(\lambda_2(d,m_2)\mid m_2)\text{cs}(\lambda_2(d,m_2)\mid m_2)\\
&\nonumber\\
&4 \beta^2 b/\eta-4\beta^2 m_2b/\eta+\frac{2\alpha^2}{\eta}\omega+\frac{2\gamma^2}{\eta} \omega \nonumber\\
&+\frac{2\beta}{\eta}[\text{cs}(\lambda_1(d,m_2)\mid m_2)\text{dn}(\lambda_1(d,m_2)\mid m_2)\nonumber\\
&-\text{cs}(\lambda_2(d,m_2)\mid m_2)\text{dn}(\lambda_2(d,m_2)\mid m_2)]\nonumber\\
&-\frac{2\alpha}{\eta}E(\alpha\omega\mid m_1)-\frac{2\gamma}{\eta}E(\gamma \omega\mid m_3)\nonumber\\
&-\frac{2\beta}{\eta}[-E(\lambda_1(d,m_2)\mid m_2)+E(\lambda_2(d,m_2)\mid m_2)]=1
\end{eqnarray}
where the same notations as in the attractive case has been used. Here the five unknown variables are $m_1$, $m_2$, $m_3$, $d$, and $\mu$; $\alpha$, $\beta$ and $\gamma$ are functions of the elliptic parameters and the chemical potential $\mu$. A similar set of equations is obtained for the solutions that has nodes inside the barrier.

\begin{figure}
\begin{center}
\epsfig{figure=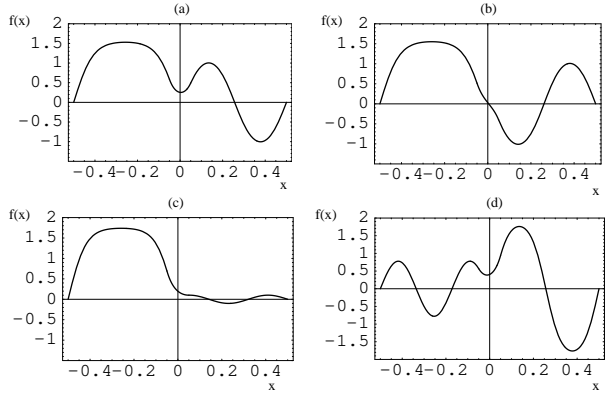,width=8.2cm}
\end{center}
\caption{Shown are the first four one-node, two-node and three-node symmetry breaking states for repulsive nonlinearity. The barrier walls are at x=$\pm$ 0.05}
\label{fig:repsb}
\end{figure}

\begin{figure}
\begin{center}
\epsfig{figure=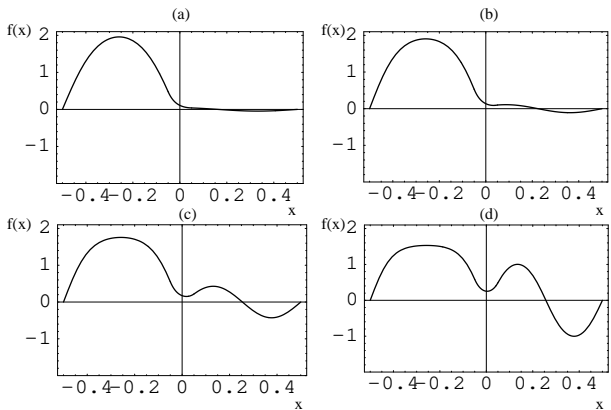,width=8.2cm}
\end{center}
\caption{Symmetry breaking repulsive ground state as a function of nonlinearity . The barrier walls are at x=$\pm$ 0.05. (a) $\eta$=15, (b) $\eta$=30, (c) $\eta$=50, (d) $\eta$=100}
\label{fig:repeta}
\end{figure}

The first four states are plotted in Fig.~\ref{fig:repsb} for a nonlinearity of $\eta=100$ and the same well dimension as the repulsive symmetry preserving case, $2a=1$, $2b=1/10$ and $V_o=1000$. Table~\ref{table:sb} shows the solution parameters for Fig.6. Again the solutions can be classified as one-node, two-node, three-node symmetry breaking states. For repulsive condensates the asymmetric ground state has a much higher energy and is in fact the second excited state of the double well. Note that for the two two-node solutions keeping one node inside the barrier and another outise the barrier is energetically more favorable than having two nodes outside the barrier. In Fig.~\ref{fig:repeta} we show the symmetry breaking ground state as we change the nonlinearity. It evolves from being almost localized for small nonlinearity to having three distinct density lumps for high enough nonlinearity.           

\section{Schr\"odinger Cat State of BEC in a double well}
\label{sec:cat} 

As was shown in the previous section, the mean field ground state of attractive condensate and some of the excited states of both attractive and repulsive condensate are symmetry breaking states. For the symmetry breaking localized states such as the attractive ground state, coherent quantum tunneling between the degenerate states removes the degeneracy and forms a superposition of the mean field states. Such localized superposition states of the form $\phi_{left}^N \pm \phi_{right}^N$ are Schr\"odinger cat states. On the other hand, the usual even and odd symmetry preserving delocalized states of the form $\Psi^N=(\phi_{left} \pm \phi_{right})^N$ are not traditional Schr\"odinger cat states. For the cat states, tunneling splitting is exponentially small in the N-body wave function overlap. In the following we find the zero order two-configuration mean field cat state tunneling splitting starting with the N-particle Hamiltonian with pseudopotential interaction. It has not gone unnoticed that the ground state of the attractive condensate is cat-like~\cite{ho1,zhou1}. Cirac {\itshape et al.}~\cite{cirac1} have studied the ground state of Josephson-coupled two-species condensates which has similarities with condensate in a double well. In the next section we deliberate on the experimental realization of cat states of BEC in a double well.      

\subsection{Schr\"odinger cat state tunneling splitting}
\label{catderivation}

The N-body Hamiltonian for a system of N weakly interacting identical bosons each of mass $m$ in an external potential $V_{ext}$ is

\begin{equation}
H_N=\sum_{i=1}^{N}\left(-\frac{\hbar^2}{2m}\nabla_i^2+V_{ext}({\bf r}_i)\right)+ 1/2 \sum_{i\neq j}V({\bf r}_i,{\bf r}_j),
\label{eqn:Nhamiltonian}
\end{equation}

Here $V({\bf r}_i,{\bf r}_j)=g\delta({\bf r}_i-{\bf r}_j)$ is the Fermi `contact' pseudo-potential, and $g=\frac{4\pi a_s \hbar^2}{m}$ where $a_s$ is the s-wave scattering length characterizing the binary atomic collisions.   

For a fully condensed Bose condensate the N-body wavefunction can be written as a symmetric product of single-particle wave functions
\begin{equation}
\Psi_N({\bf r}_1,{\bf r}_2,\ldots,{\bf r}_N)=\phi({\bf r}_1)\phi({\bf r}_2)\ldots\phi({\bf r}_N)\equiv \phi^N
\label{eqn:Nwavefnc}
\end{equation}
where $\phi({\bf r}_i)$'s are the single particle mean field wavefunctions normalized to unity $\int d{\bf r}|\phi|^2=1$.  

The expectation value gives us the N particle mean field energy
\begin{equation}
\label{eqn:sameoverlap1}
\langle\phi^N|H_N|\phi^N\rangle=\mu N -\frac{N(N+1)}{2}g\int d{\bf r}|\phi|^4\\
\end{equation}
where $\mu$ is the chemical potential.
We can generalize these to the left and right localized GP solutions in a double well,  which in the above equatin would correspond to replacing $\phi$ by $\phi_L$ and $\phi_R$. The expectation value with respect to the left and right localized states contains overlap integrals which no longer vanishes because of their non-orthogonality - 
\begin{eqnarray}
\label{eqn:overlap}
\langle\phi_L^N|H_N|\phi_R^N\rangle= \frac{N(N-1)}{2}g\int d{\bf r}(\phi_L^*)^2 \phi_R^2 \left(\Lambda \right)^{N-2}\nonumber\\
-gN^2\int d{\bf r} \phi_L^*|\phi_R|^2 \phi_R \left(\Lambda\right)^{N-1}+\mu N\left(\Lambda\right)^N
\end{eqnarray}
where $\Lambda=\int d{\bf r} \phi_L^*\phi_R$ is the overlap integral. The even and odd combinations of the left and right localized solutions $\phi_{left}^N \pm \phi_{right}^N$ are a two configuration model for Schr\"odinger cat superposition states. Taking $\phi_L$ and $\phi_R$ to be real, the expectation value of the energy at this simplest level of approximation is
\begin{eqnarray}
E_{S,A}=\frac{\langle\phi_L^N|H_N|\phi_L^N\rangle \pm \langle\phi_L^N|H_N|\phi_R^N\rangle}{1 \pm \langle\phi_L|\phi_R\rangle^N}
\label{eqn:energy}
\end{eqnarray}
Although this equation is identical in appearance with Eq.(20) of Cirac {\itshape et al.}~\cite{cirac1}, our use and inclusion of the exact mean field effect on the fully localized left and right well solutions differ from their treatment of spinor condensates.      
The tunneling splitting is the difference in antisymmetric and symmetric energy
\begin{eqnarray}
\Delta E=E_A-E_S
\label{eqn:splitting1}
\end{eqnarray}
For the case when the overlap is extrememly small and for a large number of particles the normalization factor in the denominator can be ignored and the splitting can be written as 
\begin{eqnarray}
\Delta E \approx -2 \mu N (\Lambda)^N +2gN^2 \int d{\bf r} \phi_L |\phi_R|^2\phi_R (\Lambda)^{N-1}\nonumber\\
-N(N-1)g\int d{\bf r} \phi_L^2 \phi_R^2 (\Lambda)^{N-2}
\label{eqn:splitting2}
\end{eqnarray}
This shows explicitly how the cat state tunneling splitting depends on the overlap of the localized single particle mean field wave functions. However, in our calculations we find the exact splitting by use of Eq.~(\ref{eqn:splitting1}) since physically realizable splitting can only be generated for a significant overlap such that we cannot completely ignore the the $\Lambda^N$ term in the denominator. Since $\Lambda$ is always less than 1 the splitting is exponentially small in the wavefunction overlap.

We use the solution of the one-dimensional GP equation Eq.~(\ref{eqn:nls}) to find the tunneling splitting and its dependence on other quantities. The conversion factor to get the energy from a dimensionless quantity is $\hbar^2/(2m\,l^2)$~\cite{kmahmud1}, where $l$ is the length of the box. To find the splittings in one dimension, the coupling constant `$g$' should be replaced by the dimensionless effective one-dimensional coupling constant $g_{eff}$. The dimensionless nonlinearity $\eta$ of the NLSE Eq.~(\ref{eqn:nls}) is related to $g_{eff}$ by the realtionship $\eta=g_{eff} N$, where N is the total number of particles. For experimental purposes where a condensate is three dimensional or can be quasi-one dimensional the effective coupling constant $g_{eff}$ depends on the transverse dimensions of the trap, the species of atoms (whether attractive or repulsive) and the total number of particles in a nonlinear and nontrivial way. Even without knowing the exact $g_{eff}$ for realistic three dimensional condensate we can explore the dependence of the tunneling splittings on the number of particles N and on the effective coupling constant. The relationship between the effective coupling constant $g_{eff}$ and the transverse dimensions of realistic double well traps that will give the correct experimental predictions is under investigation.               

\subsection{Discussions}
\label{}

\begin{figure}
\begin{center}
\epsfig{figure=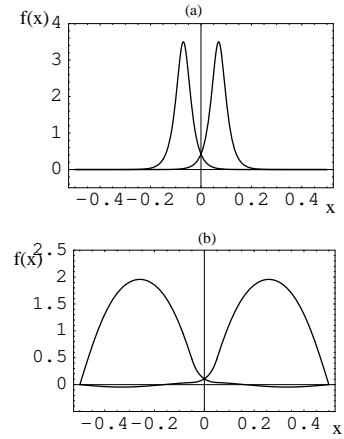,width=4.5cm}
\end{center}
\caption{A pair of symmetry breaking solutions that produces the cat states: (a) attractive ground state. (b) repulsive excited state }
\label{fig:cat}
\end{figure}

\begin{figure}
\begin{center}
\epsfig{figure=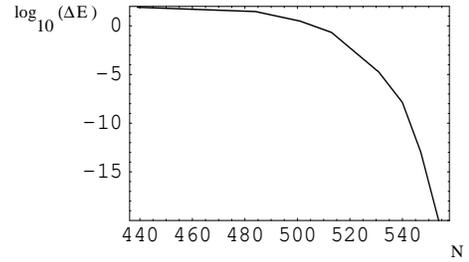,width=6cm}
\end{center}
\caption{Cat State tunneling splitting as described in the two configuration model: it shows the exponential dependence of the splitting on the number of particles for a fixed coupling constant. Energy is in frequency units of Hertz. }
\label{fig:catsplitN}
\end{figure}

\begin{figure}
\begin{center}
\epsfig{figure=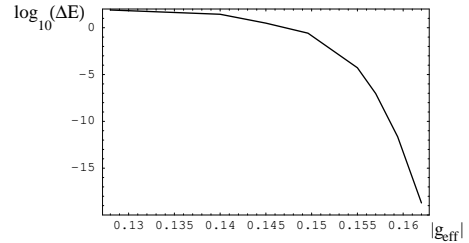,width=6cm}
\end{center}
\caption{Cat State tunneling splitting as described in the two configuration model: it shows the exponential dependence of the splitting on the effective coupling constant $|g_{eff}|$. Energy is in frequency units of Hertz. }
\label{fig:catsplitg}
\end{figure}

\begin{figure}
\begin{center}
\epsfig{figure=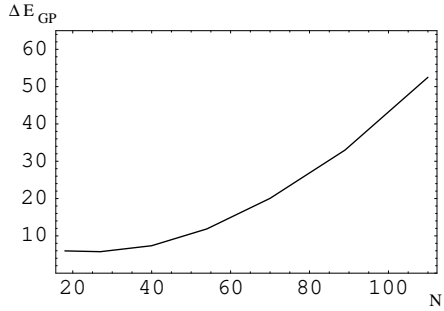,width=6cm}
\end{center}
\caption{GP single particle energy splitting between the lowest antisymmetric and symmetric state of an attractive condensate. The splitting of mean field delocalized states slowly increases with particle number, and this runs in a direction opposite to that of the Cat state tunneling splitting. Energy is in frequency units of Hertz}
\label{fig:gpsplit}
\end{figure}

Pairs of symmetry breaking mean field states in a double well are shown in Fig.~\ref{fig:cat}, coherent tunneling between these will produce a cat state. Experimentally such macroscopic cat states could be observed by starting with a localized attractive condensate in the lower well of an asymmetric double well potential, and then varying the symmetry of the two wells. In Fig.~\ref{fig:catsplitN} we show the log of tunneling splitting for a condensate of $^{7}$Li as a function of particle number for a double well of dimensions 12.5 $\mu$ separated by 75 $\mu$ in a box width of 100 $\mu$ and barrier height of $V_o=133$. A constant effective coupling constant of $g_{eff}=-0.145$ has been assumed. In Fig.~\ref{fig:catsplitg} we show the log of tunneling splitting in the same well as a function of $|g_{eff}|$ for a fixed number of particles - in this case for 500 particles. For a cat state, with the addition of more and more particles the single particle overlap becomes smaller, and the tunneling splitting becomes vanishingly small due to its exponential dependence on the overlap and the number of particles. Fig.~\ref{fig:gpsplit} shows the GP single particle tunneling splitting between the attractive antisymmetric and symmetric state for $g_{eff}=-0.911$ which sharply contrasts with the cat state tunneling splitting.

For an example of a cat state, for $N=440$ and $g_{eff}=-0.145$ the peaks of the degenerate states are asymmetrically placed on top of the barrier and the separation of the peaks is 1.5 $\mu$ , the tunneling splitting 48 Hz and the tunneling time 21 ms which are within the experimental range of detection. For higher peak separations the overlap is small and the splitting becomes negligble. An optimal cat state with gaussian barriers as is often used in experiments where the peaks are well separated and the splitting is within the range of detection should be attainable with externally tuning the coupling constant through Feshbach resonance~\cite{wieman1}. The number of particles in our study is limited to the order of hundred atoms which is within the range of stability of attractive condensates~\cite{hulet2} such as $^{7}$Li or $^{85}$Rb. Changing the scattering length by Feshbach resonance will allow stable attractive condensates to be prepared with several thousand atoms~\cite{brsolnature}. For a repulsive condensate, cat states may also be prepared making use of the excited localized condensate which must be tuned to the right regime to get a well localized condensate as shown in Fig.~\ref{fig:repeta}(a).

\section{Conclusion}
\label{sec:summary}

We have presented the stationary states of nonlinear Schr\"odinger equation in one dimension for a symmetric double square well potential for both attractive and repulsive nonlinearity. In addition to the symmetry preserving even and odd states, we find analytic expressions for symmetry breaking states that have different numbers and sizes of density lumps in the two wells. For attractive condensates these provide the analytic solutions of the stationary bright soliton trains in a double well. Symmetry breaking states do not preserve the even and odd parity of the N-particle many-body Hamiltonian. Finding such analytical solution of continous GP equation puts the self trapping states as found numerically~\cite{presilla1}, in the `two-state' tunneling models~\cite{smerzi1,milburn1,kivshar1,cirac1}, and in the discrete nonlinear Schr\"odinger equation~\cite{eilbeck} on an exact footing. Such unique symmetry breaking states, which are not possible for a linear Scr\"odinger equation, results from the nonlinearity introduced by the mean field approximation.
         
The superposition of mean field localized states of the form $\phi_{left}^N \pm \phi_{right}^N$ are Schr\"odinger cat states that arise due to coherent tunneling between the two degenerate states strongly localized in two different wells. Attractive condensate in the ground state or repulsive condensate in its symmetry breaking excited state can be used to produce such cat states. In a zero order two configuration model the splitting is exponentially small in the N-body wavefunction overlap. Tailoring the width and barrier height of a double well and with adequate number of particles in the trap to give the optimal splitting, macroscopic superposition states should be attainable with current BEC technology.  
 
The use of mean field picture in describing BEC fully delocalized in a double well is valid only when the condensate in the two wells are fully coherent. For sufficiently low tunneling, condensate in a double well cannot maintain its coherence and therefore mean field analysis of a fully coherent condensate as was presumed here is not adequate. Such fragmented condensate with number sqeezed configurations can only be treated using theories which go beyond mean field theory. However, the availability of the mean field analytic solutions as presented in this paper provides the zeroth order nonlinear wavefunctions needed to include important and large mean field effects in models which treat fragmentated condensates.


\acknowledgements

We wish to thank Lincoln Carr and Bernard Deconinck for discussions and Joachim Brand for computational support. Initial phases of this work was supported by NSF Chemistry and Physics.


\begin{references}
\bibitem[*]{byline} to whom correspondence should be addressed.
\end{references}

%

\begin{table}
\caption{Limits of Jacobian elliptic functions and integrals.  The first two sn and cn are periodic solutions in the well while dn, cn, ds, and cs are solutins in the barrier region.  $4 K(m)$ is the periodicity and the elliptic integrals $K(m)$ and $E(m)$ both play a role in the system of equations which describe the solutions.}
\label{table:jacobian}
\begin{tabular}{ccc}
  & $m=0$ & $m=1$ \\
\tableline
$\text{sn}(u \mid m)$ & $\sin (u)$  & $\text{tanh}(u)$ \\
$\text{cn}(u \mid m)$ & $\cos (u)$  & $\text{sech}(u)$ \\
$\text{dn}(u \mid m)$ & 1         & $\text{sech}(u)$ \\
$\text{ds}(u \mid m)$ & $\text{csc}(u)$  & $\text{csch}(u)$ \\
$\text{cs}(u \mid m)$ & $\text{cot}(u)$  & $\text{csch}(u)$ \\
$K(m)$                  & $\pi/2$   & $\infty$       \\
$E(m)$                  & $\pi/2$   & 1              \\
\end{tabular}
\end{table}

\begin{table}
\caption{Solutions parameters for symmetry preserving states of attractive and repulsive nonlinearity for Fig.2 and Fig.3. The numbers shown are of sufficient precision as initial estimate to be used in the numerical solution of the nonlinear equations of section II. However, as $m\rightarrow 1$ use of high precision arithmetic is required.}
\label{table:sp}
\begin{tabular}{cccccc}
  & $m$ & $m_2$ & $k$ & $\mu$\\
\tableline
$Fig.2a$   & $0.9684$  & $0.9959$ & $13.25$ & $-164.42$\\
$Fig.2b$       & $0.9758$  & $0.9935$ & $13.04$ & $-161.90$\\
$Fig.2c$       & $0.6352$  & $0.9298$ & $12.47$ & $-42.03$ \\
$Fig.2d$        & $0.4763$  & $0.7426$ & $15.36$ & $11.18$\\
$Fig.3a$   & $0.8539$  & $0.9976$ & $9.88$ & $181.06$\\
$Fig.3b$       & $0.8514$  & $0.9977$ & $9.98$ & $184.51$\\
$Fig.3c$      & $0.4338$  & $0.9912$ & $14.79$ & $313.75$ \\
$Fig.3d$     & $0.4313$  & $0.9909$ & $15.00$ & $322.24$\\
\end{tabular}
\end{table}

\begin{table}
\caption{Solutions parameters for symmetry breaking states of attractive and repulsive nonlinearity for Fig.4 and Fig.6. The numbers shown are of sufficient precision as initial estimate to be used in the numerical solution of the nonlinear equations of section II. However, as $m\rightarrow 1$ use of high precision arithmetic is required.}
\label{table:sb}
\begin{tabular}{cccccc}
  & $m_1$ & $m_2$ & $m_3$ & $d$ & $\mu$\\
\tableline
$Fig.4a$      & $0.9999$  & $1-10^{-8}$ & $1-10^{-16}$ & $-0.0680$ & $-625.27$\\
$Fig.4b$           & $1-10^{-8}$  & $0.9999$ & $0.7640$ & $0.0618$ & $-174.10$\\
$Fig.4c$           & $0.8257$  & $0.9994$ & $0.6171$ & $0.0219$ & $-62.829$ \\
$Fig.4d$           & $0.8401$  & $0.9992$ & $0.6177$ & $0.0184$ & $-62.820$\\
$Fig.6a$       & $0.9273$  & $0.9958$ & $0.2612$ & $-0.0035$ & $243.16$\\
$Fig.6b$           & $0.9273$  & $0.9959$ & $0.2529$ & $-0.0043$ & $248.95$\\
$Fig.6c$           & $0.9612$  & $0.9992$ & $0.0016$ & $-0.0491$ & $308.14$ \\
$Fig.6d$           & $0.0787$  & $0.9876$ & $0.6012$ & $0.0122$ & $412.30$\\
\end{tabular}
\end{table}

\end{document}